# Quasi-classical Trajectory Calculations on a Two-state Potential Energy Surface Including Nonadiabatic Coupling Terms as Friction for $D^+ + H_2$ Collisions

Soumya Mukherjee,[1,2] Swagato Saha,[1] Sandip Ghosh,[3] Satrajit Adhikari,[1] Narayanasami Sathyamurthy[4,*] and Michael Baer[5]

[1)] School of Chemical Sciences, Indian Association for the Cultivation of Science, Kolkata700032, India.

[2)] Department of Chemistry, Maulana Azad National Institute of Technology, Bhopal 462003, India.

[3)] Department of Chemical Sciences, Indian Institute of Science Education and Research Kolkata, Mohanpur, Nadia 741246 India

[4)] Department of Chemical Sciences, Indian Institute of Science Education and Research Mohali, Sector 81, SAS Nagar, Manauli 140306 India

[5)] The Fritz Haber Center for Molecular Dynamics, The Hebrew University of Jerusalem, Jerusalem 91904, Israel.

## Abstract

Akin to the traditional quasi-classical trajectory method for investigating the dynamics on a single adiabatic potential energy surface for an elementary chemical reaction, we carry out the dynamics on a 2-state *ab initio* potential energy surface including nonadiabatic coupling terms as friction terms for $D^+ + H_2$ collisions. It is shown that the resulting dynamics correctly accounts for nonreactive charge transfer, reactive non charge transfer and reactive charge transfer processes. In addition, it leads to the formation of triatomic $DH_2^+$ species as well.

*Email: nsathyamurthy@gmail.com

# 1. Introduction

Investigating the dynamics of a chemical event by solving the time-dependent or time-independent Schrödinger equation for the nuclei and the electrons contained in the system is a formidable challenge. Often one resorts to the Born-Oppenheimer (BO) approximation and computes the potential energy surface (PES) and carries out the quantum dynamics to predict and interpret the experimentally measured differential/integral reaction cross sections over a limited range of energy and rate coefficients over a range of temperature, particularly for 3- and 4-atom systems [1-4]. When the BO approximation fails [5-8], the quantum mechanical solution makes use of the non-adiabatic coupling between PESs explicitly [For example, see ref. 9-19]. While the quasi-classical trajectory method [20-23] has been successful in describing the dynamics on a single (adiabatic) PES for many systems, trajectory surface hopping (TSH) and a variety of semiclassical and related methods [24-42] have been used to consider switching between surfaces.

Recently, we had proposed [43] that the classical equations of motion on the ground state PES could be solved by including the non-adiabatic coupling terms (NACTs) as equivalent to friction and demonstrated the utility of such an approach in computing some of the measurables for ($D^+$, $H_2$) collisions. In addition, this method allows us to examine the possibility of formation of the triatomic species that is of astrophysical interest [44,45].

The general understanding [46] is that $H_2^+$, produced through cosmic ray (c.r.) ionization of $H_2$ reacts with the abundant $H_2$ in dense clouds and diffuse clouds to produce $H_3^+$:

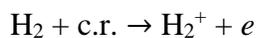

$H_2 + c.r. \rightarrow H_2^+ + e$

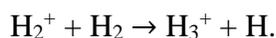

$H_2^+ + H_2 \rightarrow H_3^+ + H.$

Once formed, $H_3^+$ donates a proton readily to a variety of atoms and molecules present in the interstellar environment as discussed elsewhere [47]. As we had argued earlier [43], $H_3^+$ and its isotopomer $H_2D^+$ could be formed readily through nonadiabatic interaction between $H^+(D^+)$ and $H_2$.

Since the collision of $D^+$ with $H_2$ has been investigated experimentally [48-53] as well as theoretically using time-independent [54,55] and time-dependent [56-60] quantum mechanical methods, and trajectory surface hopping methods [34], we undertake to study the same dynamics using the quasiclassical trajectory method on a 2-state ab initio PES treating the NACTs as friction. This helps in demonstrating the possibility of trapping (triatomic species formation). Comparison of the computed cross section results for other dynamical processes in the system with the available experimental and theoretical results helps in validating the approach.

In an earlier study[61] we had examined the possibility of the NACTs acting as a friction force slowing down the dynamics of ($H_2$, $H^+$) collisions in $C_{2v}$ geometry and along parallel axes close to the $C_{2v}$ geometry. For practical reasons, $H_2$ was taken to be in its equilibrium geometry. For the initial kinetic energy (KE) ranging from 0.35 eV to 500 eV, it was shown that friction could indeed slow down the dynamics and bring the molecular system to a standstill (KE = 0). A similar study [62] was carried out for (He, $H_2^+$) collisions in collinear geometry and in $C_{2v}$ geometry for a similar energy range and it was shown that the NACT terms could indeed act as friction forces and slow down the dynamics to make KE = 0.

In a follow up study [43], we solved the classical equations of motion for the ($D^+$, $H_2(v = 0, j = 0)$) collisions in three dimensions on the ab initio PES [63] for the ground electronic state of the system and computed the integral cross sections (ICSs) for the nonreactive non-charge transfer (NRNCT) process as well as the reactive non-charge transfer (RNCT) process for the collision energy ($E_{coll}$) in the range 1.7-2.5 eV. The results were shown to be in accord with the quantum dynamical results in that energy range [43]. We repeated the classical trajectory calculations on the ground electronic state PES by including the NACTs between the ground and the first excited electronic states as friction terms and computed the ICSs for the NRNCT and RNCT processes. The difference between the sums of the ICSs for the two processes computed without friction and with friction was attributed to the trapping ($DH_2^+$ formation) process. We also followed the kinetic energy, potential energy and the total energy for individual trajectories as a function of time up to 25 ps on the ground electronic PES (by including the friction terms) to confirm that the trajectories came to a halt at a total energy of -8.395 eV, close to the classical minimum of the PES corresponding to $DH_2^+$. Bond distances were used to characterize the different channels:

|$R_{H1}$-$R_D$| > 3 Å and |$R_{H2}$-$R_D$| > 3 Å (NRNCT),

|$R_{H1}$-$R_D$| or |$R_{H2}$-$R_D$| > 3 Å and |$R_{H1}$-$R_{H2}$| > 3 Å (RNCT) and

|$R_{H1}$-$R_{H2}$| ~ |$R_{H1}$-$R_D$| ~ |$R_{H2}$-$R_D$| ~ 0.873 Å ($DH_2^+$ formation),

where $R_{H1}$, $R_{H2}$ and $R_D$ refer to the position of the two H atoms and the D atom, respectively.

When friction was included, the QCT calculations yielded ICSs for NRNCT and RNCT processes slightly lower than those obtained when friction was not included. The ICS for trapping decreased from 1.2 Å² at $E_{coll}$ = 1.7 eV to 0.4 Å² at $E_{coll}$ = 2.5 eV. That study clearly demonstrated that NACTs could act as friction and slow down the trajectories on the ab initio PES to result in $DH_2^+$ formation.

In what follows we present the theoretical background going beyond the Born-Oppenheimer approximation and formulate the classical trajectory methodology for a 2-state PES. Since the relevant ab initio potential energy surfaces and the NACTs have been reported earlier, we make use of them in the present study. We have computed the integral cross section values for NRNCT and RNCT processes in the ground electronic state, and NRCT and RCT processes in the first excited electronic state for the maximum impact parameter ($b_{max}$) = 3 Å for $E_{coll}$ = 1.7 eV, 2.5 eV and 5.0 eV by using the NACTs as friction coefficients in classical trajectory calculations (using a cut off of 0.05 Å at the origin of the Cartesian coordinates), initiating the trajectories from the ground adiabatic surface of $DH_2^+$ asymptotically corresponding to $D^+$ + $H_2$ ($v$ = 0, $j$ = 0).

## 2. Theoretical background

### 2.1 A quantum Mechanical Approach

Considering the electronic eigenfunctions ($\zeta_j$) at the nuclear coordinate $s$, we obtain the following set of Born-Oppenheimer-Huang (BOH) equations [64-66] for the nuclear motion:

$$-\frac{\hbar^2}{2m}\nabla^2\psi_k + (u_k - E)\psi_k - \frac{\hbar^2}{2m}\sum_{j=1}^{N}\left(2\tau_{kj}\cdot\nabla + \tau_{kj}^{(2)}\right)\psi_j = 0; k = \{1, N\}, \quad (1)$$

where $\tau_{jk}(s)$ is the first order (vector) non-adiabatic coupling term (NACT) defined as:

$$\tau_{jk}(s) = \langle\zeta_j|\nabla\zeta_k\rangle, \quad (2)$$

and $\tau_{jk}^{(2)}(s)$ is the second order (scalar) NACT defined as:

$$\tau_{jk}^{(2)}(s) = \langle \zeta_j | \nabla^2 \zeta_k \rangle. \tag{3}$$

The symbols $\nabla, E, \mathbf{u}, \hbar$ and $m$ have their usual meaning [43]. For a system of real electronic wave functions $\boldsymbol{\tau}(s)$ is an anti-symmetric (vector) matrix.

Eq. (1) can also be written in a matrix form as follows [6a]:

$$-\frac{\hbar^2}{2m}\nabla^2 \boldsymbol{\Psi} + (\mathbf{u} - E)\boldsymbol{\Psi} - \frac{\hbar^2}{2m}\left(2\boldsymbol{\tau} \cdot \nabla + \boldsymbol{\tau}^{(2)}\right)\boldsymbol{\Psi} = \mathbf{0}, \tag{4}$$

where the dot designates the scalar product, $\boldsymbol{\Psi}(\mathbf{s})$ is a column vector that contains the above-mentioned nuclear functions $\{\psi_j(\mathbf{s}), j = [1, N]\}$ and $\mathbf{u}(\mathbf{s})$ is a diagonal matrix which contains the adiabatic potential energy surfaces. This equation is valid for any group of states.

Making use of the relation between $\boldsymbol{\tau}$ and $\boldsymbol{\tau}^{(2)}$ [6a], Eq. (4) can be rewritten as:

$$-\frac{\hbar^2}{2m}\nabla^2 \boldsymbol{\Psi} + \left(\mathbf{u} - \frac{\hbar^2}{2m}\boldsymbol{\tau}^2 - E\right)\boldsymbol{\Psi} - \frac{\hbar^2}{2m}(2\boldsymbol{\tau} \cdot \nabla + \nabla\boldsymbol{\tau})\boldsymbol{\Psi} = \mathbf{0}, \tag{5}$$

Eq. (5) can also be written in a more compact way [6a]

$$-\frac{\hbar^2}{2m}(\nabla + \boldsymbol{\tau})^2 \boldsymbol{\Psi} + (\mathbf{u} - E)\boldsymbol{\Psi} = \mathbf{0}. \tag{6}$$

Eq. (6) is the nuclear BOH differential equation within the **Adiabatic** framework for a group of states that forms a Hilbert space [6a]. In this sense Eq. (6) differs from Eq. (4), which is valid for any group of states.

Since the NACTs are known to possess singularities, solving Eq. (6) as such becomes impossible unless the singularities are eliminated. This is often accomplished by **Diabatization[6]** but uneasiness is involved in applying this process because the original Eq. (6) does not conserve energy whereas the **diabatic** differential equations do. Still, the only way to study quantum mechanically molecular processes (without approximations) as of now is by solving the **diabatic** equations.

## 2.2 A classical approach

As is noticed, Eq. (6) contains two kinds of forces: (i) forces which have their origin in the PESs formed by electrostatic interactions and (ii) forces which have their origin in the NACTs and are seen to be of a different kind due to the way they show up in the equation, namely, being associated with velocities and therefore are known as Forces of Friction. To explore this possibility, it was suggested to apply Classical Mechanics (two studies were performed [61, 62] and both were found to justify the idea that indeed, NACTs could be considered as a source for the force of friction). The advantage in applying Classical Mechanics for this purpose is that solving the classical equations of motion can be done also for singular NACTs (see for instance Ref. 43). Classical Equations of motion containing forces of Friction are discussed in various textbooks (for example, see ref. 67, 68) and the typical case is an equation of motion for the free mode vibrations affected by force of friction proportional to the velocity.

The basic classical equation of motion (in one dimension), which contains the friction term, $\beta \dot{x}$, takes the form[67,68]

$$m\ddot{x} + \beta \dot{x} + \frac{dV(x)}{dx} = 0 \qquad (7)$$

where the terms $\dot{x}$ and $\ddot{x}$ stand for $(dx/dt)$ and $(d^2x/dt^2)$, respectively, $V(x)$ is the potential energy and $\beta$ is the corresponding (x-component) friction coefficient assumed to be positive throughout any numerical treatment.

Eq. (7) can be written also as an energy equation [61,62]:

$$\left[\frac{1}{2m}(m\dot{x} + \beta x)^2 + V(x)\right] = E + G(x, \dot{x}) \qquad (8)$$

where $E$ is a constant (and is the total energy in case $\beta = 0$) and $G(x, \dot{x})$ is given in the form

$$G(x, \dot{x}) = \int x \left[-\frac{1}{m}\beta \frac{dV}{dx} + \frac{d\beta}{dt}\left(\frac{1}{m}\beta x + \dot{x}\right)\right] dt. \qquad (9)$$

The similarity between the quantum mechanical equation (6) and the classical equation of motion (8) is not obvious since the former is a matrix-equation that has to be applied to a group of coupled equations whereas the latter is an equation for a single isolated state. Still such a similarity can be envisaged by applying a derivation of a **modified** BO approximation as was carried out by Baer and Englman [69, 6c] for two coupled equations (in a single coordinate). This approximation yields in case of low enough energies, the following single equation:

$$\frac{1}{2m}\left(\frac{\hbar}{i}\frac{\partial}{\partial x} + i\hbar\tau_{x12}(x)\right)^2 \Psi(x) + (u_1(x) - E)\Psi(x) = 0 \qquad (10)$$

It is important to emphasize that the solution of Eq. (10) is reliable for all the conditions/situations envisaged in the present study. (The relevance of this approximation was tested extensively in Ref. 70 – see Tables (II)-(V)).

Eqs. (8) and (10) are now in a form to apply **Schrödinger's correspondence principle** to reveal the connection between the quantum mechanical $x$-component of $\tau_{x12}(x)$ – see Eq. (10) – and the corresponding classical friction coefficient, $\beta(x)$ (which is known to be responsible for the classical dissipative force (or friction force) [61, 62]. Therefore, it follows that

$$\hbar\tau_{x12}(x|s) = \beta(x)x \;\Rightarrow\; \beta(x) = \frac{\hbar\tau_{x12}(x|s)}{x}. \qquad (11)$$

Such a relation holds for other Cartesian coordinates $y$ and $z$ as well, where is required. Here, it must be emphasized that the various NACTs are pure quantum mechanical entities and are being used as classical friction components while solving classical equations of motion for the molecular systems. Consequently, the whole treatment becomes semi-classical.

According to Classical mechanics (see for instance References 67 and 68) the friction coefficient must be positive (or zero) in all situations so this should apply also for its expression on the RHS. However, according to quantum mechanics the NACTs might be either negative or positive. Moreover, an analogous situation holds also for the denominators: the coordinates ($x$, $y$, $z$), which might be either positive or negative. Thus, according to the **uncertainty principle,** both negative and positive ratios are acceptable but according to classical mechanics only the positive ratios are acceptable (negative ratios accelerate the atomic particles which according to classical mechanics is unacceptable). Thus, one possibility is to assume the friction term to be positive "all the way", namely.

$$\beta_{qj}(q|s) = \hbar \left| \tau_{qjj+1}(q|s)/q \right|, \quad q = x, y, z. \tag{I}$$

In what follows it will be termed as the **Classical Friction**.

Another possibility which is introduced under the influence of quantum mechanics is:

$$\beta_{qj}(q|s) = \begin{cases} \hbar \tau_{qjj+1}(q|s)/q & \text{when} \quad \tau_{qjj+1}(q|s)/q \geq 0 \\ 0 & \text{when} \quad \tau_{qjj+1}(q|s)/q \leq 0 \end{cases} \tag{II}$$

This possibility is reminiscent of **Schrodinger's cat** which according to its definition maybe either **alive or dead**. In what follows it will be termed as **Semi-classical Friction.**

The third possibility is to assume the friction term to be as it is:

$$\beta_{qj}(q|s) = \hbar \tau_{qjj+1}(q|s)/q \tag{III}$$

This possibility is justified by the fact that the NACTs are singular and therefore capable not only to slow down the interacting atomic/ionic particles but eventually also to accelerate them. This type of friction will be termed as **Quantum Friction**.

There is a good reason to apply the quantum version – a reason associated with the (artificial) singularities which are encountered once a particle crosses a $q$-axis ($q = x, y, z$). Assuming that happens while $\beta(q)$ is positive the large acceleration is soon abolished once the particle crossed the $q$-axis (causing $\beta(q)$ to become negative) and consequently the particle is slowed down at a similar rate. Only within this version the effect of the artificial singularity is, essentially, annihilated.

It is important to emphasize that just like the adiabatic BOH equations (eq. (1) or Eq. (6)) do not conserve energy, the semi-classical Eq. (7) also does not conserve energy. There is one issue which plays in favour of the NACTs being associated with friction. That is, during the slowing down process it releases energy as heat − in the present case the energy would be released as light (or, better, as photons), as envisaged in the production of HeH$^+$ in (He$^+$, H$_2$) collisions (see Ref. [71] and discussion in Ref. [62]).

The dissipation process is enhanced once three or more quasi-ions are in relative proximity because, in such a situation, the NACTs attain, at certain regions in configuration space, infinite values. As a result, the fast-moving quasi-ions are continuously slowed down and in this way are forced to lose their excess energy at various rates so that the electrostatic forces may, finally, takeover.

Influence of molecular friction on reaction dynamics in condensed phases has been studied for several decades now [72]. Recently, there has been a lot of interest in electronic friction arising from non-adiabatic coupling particularly in gas-metal interactions [73-82]. Such an approach using a single potential energy surface seems successful because of a continuum of electronic states present in the metal, serving as a bath. The latter is treated as a dissipative force and a random force. However, our approach involves two electronic states, and the non-adiabatic coupling is explicitly treated as friction. While the electronic friction in gas-metal system has been shown to be applicable in the weak non-adiabatic limit, our approach is valid in two or more electronic states in the entire configuration space for a molecular system.

We have reported recently [63] the ab initio PES for the ground and the first excited electronic state of $(DH_2)^+$ along with the NACTs in three dimensions. Although the two channels $(H_2, H^+)$ and $(H_2^+, H)$ are degenerate in the limit $(H + H + H^+)$, they differ in their energies in the intermediate regions in configuration space as illustrated in Figure 1.

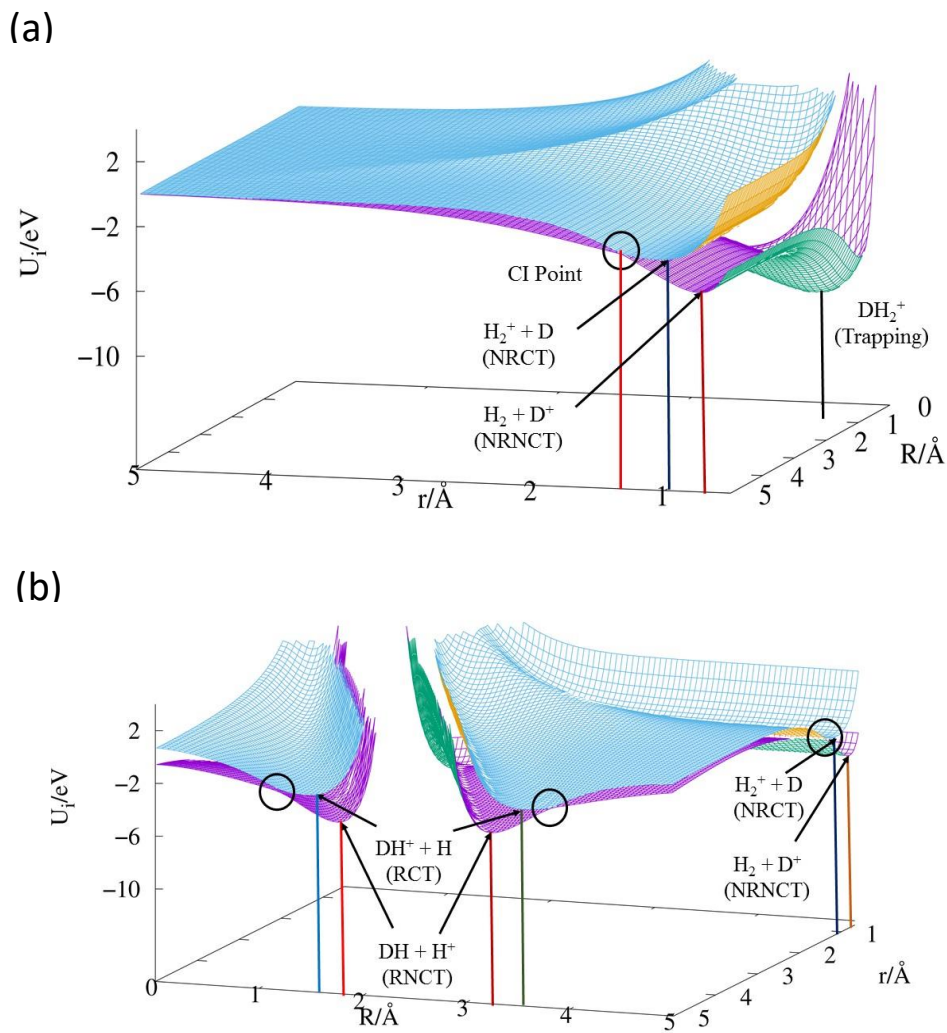

Figure 1: Potential energy surfaces for the ground and the first excited electronic state of the $DH_2^+$ system as a function of the Jacobi coordinates $r$ and $R$ for (a) $C_{2v}$ and (b) collinear geometries [63].

In the present study, we have solved the classical equations of motion by including friction terms equivalent to the NACTs. It is shown that the $(D^+, H_2)$ dynamics slows down enough to reach the bottom of the potential well. Therefore, we argue that a classical mechanical study including the NACTs would lead most likely, to the formation of $DH_2^+$ in its equilibrium geometry.

For the $(DH_2)^+$ system, a wide variety of NACTs are spread over the entire nuclear configuration space. One can observe accidental seams between the ground ($1^1A'$) and the first excited ($2^1A'$) states of $(DH_2)^+$ at the $C_{2v}$ (isosceles) as well as $C_s$ (acute) geometries over the global PESs, whereas the CIs between the first ($2^1A'$) and the second ($3^1A'$) excited states are located at $D_{3h}$ (equilateral) nuclear configurations as illustrated elsewhere [63]. Therefore, it is necessary to extend the classical dynamical calculations over the global three-dimensional (3D) PES (rather than the one-dimensional potential energy curve investigated in ref. 61) of $H_3^+$ to compute the trapping percentage in the global minimum (-9.31 eV relative to isolated ($D^+$, $H_2$) and ($D$, $H_2^+$)) as $DH_2^+$ species and other reaction attributes (reaction probabilities and cross sections) of various processes resulting from ($D^+$, $H_2$) interactions.

## 3. Classical trajectory methodology for $D^+$ + $H_2$ Collisions on the Ground and the First Excited Adiabatic Potential-Energy Surfaces

For $D^+$ + $H_2$ collisions, a trajectory is initiated with non-zero momentum corresponding to $E_{coll}$ between the reactants in the ground electronic state and a corresponding trajectory on the excited state starts with zero momentum, and the following competing processes can take place:

- Creation of Triatomic Species (trapping near the potential energy minimum of the ground electronic state):

  $D^+ + H_2 \rightarrow DH_2^+$,

- Non-Reactive Non-Charge Transfer (NRNCT):

  $D^+ + H_2 \rightarrow D^+ + H_2$,

- Reactive Non-Charge Transfer (RNCT):

  $D^+ + H_2 \rightarrow H^+ + DH$,

- Reactive Charge Transfer (RCT) (first excited state):

  $D^+ + H_2 \rightarrow DH^+ + H$,

- Non-Reactive Charge Transfer (NRCT) (first excited state):

  $D^+ + H_2 \rightarrow D + H_2^+$

We have carried out classical trajectory calculations on the ground electronic state PES as well as the 2-state PES at $E_{coll}$ = 1.7, 2.5 and 5.0 eV for the maximum impact parameter ($b_{max}$) of 3 Å by including the NACTs as friction coefficients (using a cut-off value of 0.05 Å at the origin of the Cartesian coordinates). Initialization of the trajectories is done on the ground adiabatic surface of $DH_2^+$ asymptotically representing $D^+ + H_2$ ($v = 0, j = 0$), where $v$ and $j$ represent the vibrational and rotational quantum numbers, respectively.

While there are 18 equations of motion (EOMs) for positions and momenta on a single PES (see ref. 43), 36 equations of motion are involved for the 2-state problem as given below:

$$\dot{p}_{1x_k} + \frac{\hbar}{m_k x_k} p_{1x_k} \tau_{12}^{1x_k} + \frac{dV_1(x_k, y_k, z_k)}{dx_k} = 0$$

$$\dot{p}_{2x_k} - \frac{\hbar}{m_k x_k} p_{2x_k} \tau_{12}^{2x_k} + \frac{dV_2(x_k, y_k, z_k)}{dx_k} = 0$$

$$\dot{p}_{1y_k} + \frac{\hbar}{m_k y_k} p_{1y_k} \tau_{12}^{1y_k} + \frac{dV_1(x_k, y_k, z_k)}{dy_k} = 0$$

$$\dot{p}_{2y_k} - \frac{\hbar}{m_k y_k} p_{2y_k} \tau_{12}^{2y_k} + \frac{dV_2(x_k, y_k, z_k)}{dy_k} = 0$$

$$\dot{p}_{1z_k} + \frac{\hbar}{m_k z_k} p_{1z_k} \tau_{12}^{1z_k} + \frac{dV_1(x_k, y_k, z_k)}{dz_k} = 0$$

$$\dot{p}_{2z_k} - \frac{\hbar}{m_k z_k} p_{2z_k} \tau_{12}^{2z_k} + \frac{dV_2(x_k, y_k, z_k)}{dz_k} = 0$$

$$\dot{x}_{1k} = \frac{p_{1x_k}}{m} \quad \dot{y}_{1k} = \frac{p_{1y_k}}{m} \quad \dot{z}_{1k} = \frac{p_{1z_k}}{m}$$

$$\dot{x}_{2k} = \frac{p_{2x_k}}{m} \quad \dot{y}_{2k} = \frac{p_{2y_k}}{m} \quad \dot{z}_{2k} = \frac{p_{2z_k}}{m} \qquad (13)$$

It may be noted that the classical trajectory calculations are performed for the 2-state cases involving (I) $|\tau_q/q|$ throughout the trajectory, i.e., using Landau-Lifshitz (LL) condition (Classical Friction) as well as (II) inclusion of $\tau_q/q$ in the classical EOMs, only when the product of $\tau_q$ and $1/q$ is naturally positive (Semiclassical Friction). In other words, for the second case, if the quantity $\tau_q/q$ is positive, it is included in the classical EOMs, otherwise set to zero. The sign of $\dot{q}$ is taken as it is,

i.e., no constraint is imposed on the sign of $\dot{q}$. In addition, we have also carried out the calculations without including any sign correction in the friction term. That is, choice III (quantum friction).

The cross-section ($\sigma$) values for any dynamical process are calculated using the following expression:

$$\sigma = \frac{\pi}{\kappa_{vj}^2} \sum_{v'=0}^{v'_{max}} \sum_{j'=0}^{j'_{max}} \sum_{N_{v'j'}} (l_{v'j'} + 1)^2 \left(\frac{1}{N_T}\right) \tag{14}$$

where $N_{v'j'}$ represents the number of trajectories leading to the reactant/product at a specific rovibrational state $(v', j')$. $N_T$ represents the total number of trajectories. The orbital angular momentum quantum number ($l$) takes the following form:

$$l_{v'j'} = l_{max} O_{v'j'} = \left(\frac{b_{max}\mu_{tri}v_0}{\hbar} - \frac{1}{2}\right) O_{v'j'} \tag{15}$$

where $b_{max}$, $\mu_{tri}$, $v_0$ and $O_{v'j'}$ are maximum impact parameter, triatomic center of mass, initial relative velocity and pseudo random number, respectively. The detailed formulation of the above two equations is given in the Appendix.

## 4. Results and Discussion

Since the classical trajectory methodology is well known [20-23], we only state that we computed 6000 trajectories for different initial conditions for a given $E_{coll}$ by ensuring convergence with respect to $R$. We computed initially the probability and then the cross section for trapping (formation of $DH_2^+$), NRNCT, RNCT, RCT and NRCT processes. It is important to reiterate that the RCT and NRCT processes end up on the first excited electronic state.

It Is important to point out that the earlier quantum dynamical calculations [57] considered the orbital angular momentum quantum number ($l$) up to 40 while the trajectory surface hopping (TSH) calculations [34] considered $b_{max}$ = 3 Å. Therefore, we have also used the $b_{max}$ value of 3 Å.

For illustrating the effect of friction in slowing down the trajectories and leading to trapping on the 2-state PES, plotted in Figure 2 are different interatomic distances in the ground as well as in the excited electronic state, respectively for a representative trajectory with $E_{coll}$ = 1.7 eV. It is important to point out that the $R_{H1-D}$ and $R_{H2-D}$ distances are larger for the excited state because the asymptotes are different for the excited state than for the ground state.

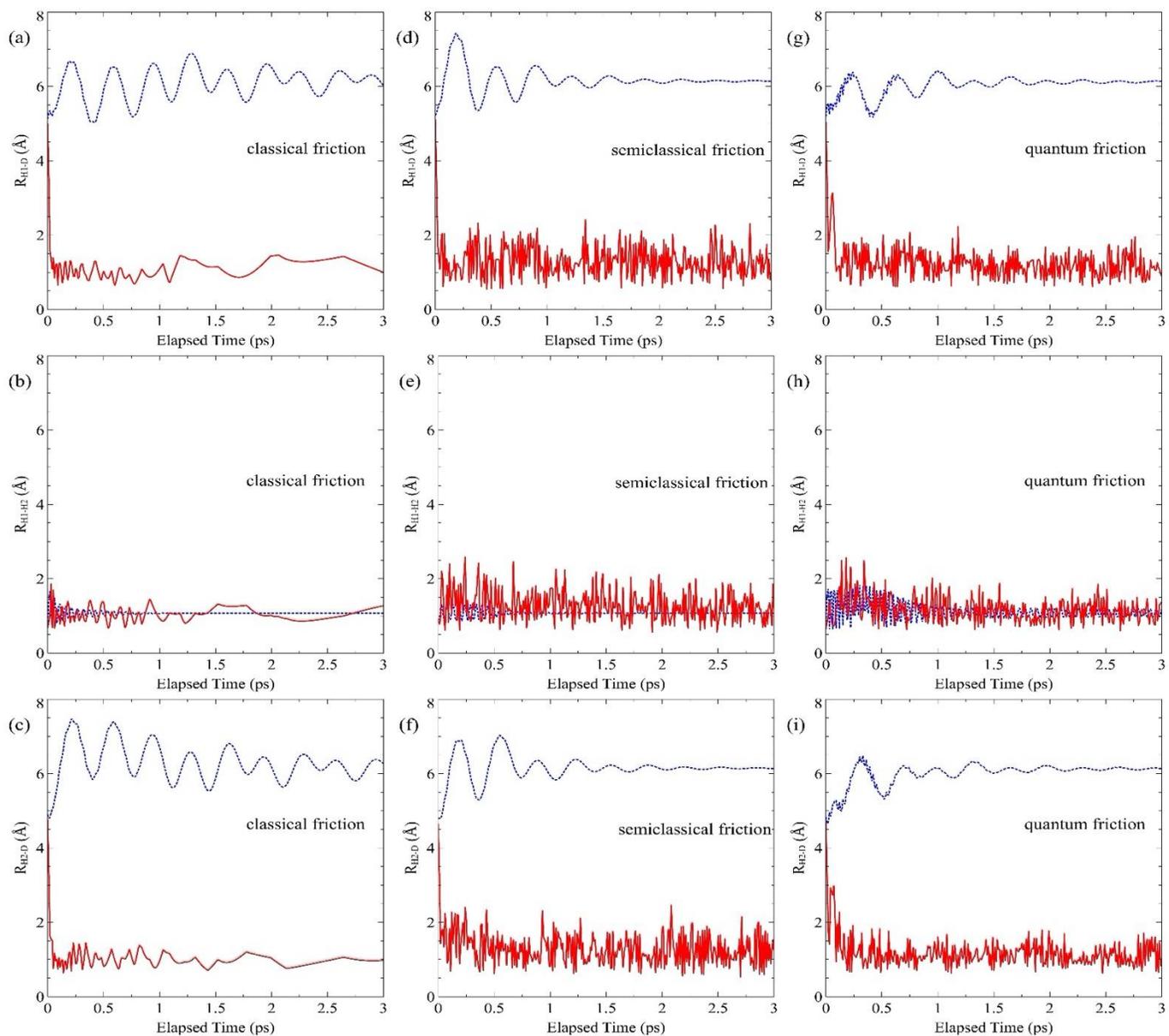

Figure 2: Variation of interatomic distances, (a,d,g) $R_{H1-D}$, (b,e,h) $R_{H1-H2}$ and (c,f,i) $R_{H2-D}$ with respect to elapsed time for a representative trajectory at $E_{coll}$ = 1.7 eV ($b_{max}$ = 3 Å) on the ground (red solid lines) and first excited (blue dotted lines) electronic state PES (using three types of friction). The trajectory is considered trapped when the ground state bond lengths are $R_{H1-D}$ = 0.88 Å, $R_{H1-H2}$ = 0.96 Å and $R_{H2-D}$ = 0.72 Å for classical friction; $R_{H1-D}$ = 1.31 Å, $R_{H1-H2}$ = 1.77 Å and $R_{H2-D}$ = 0.92 Å for semiclassical friction and $R_{H1-D}$ = 1.19 Å, $R_{H1-H2}$ = 1.20 Å and $R_{H2-D}$ = 0.77 Å for quantum friction. Excited state bond lengths correspond to $H_2^+$ + D as the asymptote.

The variation of kinetic energy, potential energy and the total energy with respect to time is illustrated in Figure 3. While the kinetic energy and the potential energy oscillate as a function of time, the total energy would have remained constant with change in time, if there was no friction. However, it is clear from Figure 3 that the total energy has decreased over a period of time, reaching the lowest energy possible in about 3 ps. Clearly, this particular trajectory is "trapped", resulting in the formation of $DH_2^+$.

To illustrate the role of friction a la NACTs in determining the collisional outcome in a 2-state calculation, we plot the *x*- and *y*-components of the friction coefficient in the ground and the excited electronic states in Figures S1 and S2 of Supporting Information, respectively. It is worth adding that the way the classical trajectory calculations are carried out, the dynamics is largely in the (*x*, *y*) plane. Therefore, the *z*-component of friction remains zero.

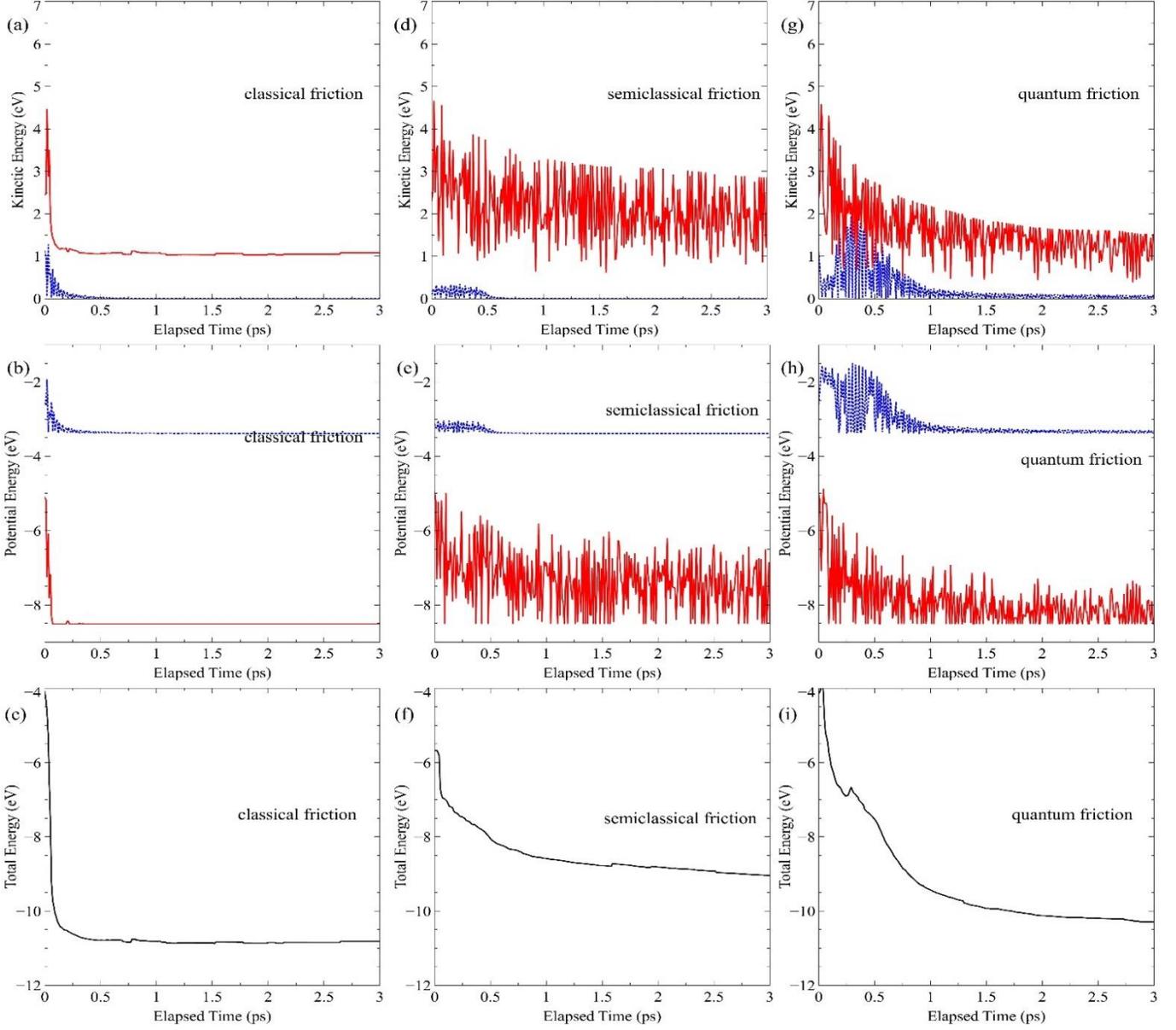

Figure 3: Variation of (a,d,g) kinetic energy, (b,e,h) potential energy and (c,f,i) total energy with time for a representative trajectory at $E_{coll}$ =1.7 eV ($b_{max}$ = 3 Å) for the ground (red solid lines) and first excited electronic (blue dotted lines) states of $DH_2^+$. The trajectory is considered trapped when the ground state potential energies ($V_{min}^T$) are -8.51 eV, -7.44 eV and -8.51 eV for classical, semiclassical and quantum friction, respectively. On the other hand, total energies are -10.45 eV, -9.71 eV and -10.61 eV, respectively.

In case of untrapped trajectories, the ground state of a specific trajectory leads to either NRNCT or RNCT (reactants/products) and at the same time, the excited state component of the same trajectory leads to either NRCT or RCT (reactants/products). In other words, the number of (NRNCT + RNCT) trajectories will be the same as the number of (NRCT + RCT) trajectories.

For $E_{coll}$ = 1.7 eV and $b_{max}$ = 3 Å, out of 6000 trajectories using classical friction, 5236 are trapped and 764 are not trapped. Among the ground state components of untrapped trajectories, 699 lead to NRNCT and 65 lead to RNCT processes. Among the excited state components, 657 lead to NRCT and 107 lead to RCT (reactants/products). In case of semiclassical friction, the numbers are 2949 (trapping), 2257 (NRNCT), 794 (RNCT), 2594 (NRCT) and 457 (RCT), respectively. If no sign correction (quantum friction) is involved, the numbers are 167 (trapping), 3741 (NRNCT), 2092 (RNCT), 3193 (NRCT) and 2640 (RCT), respectively.

In comparison, for $E_{coll}$ = 2.5 eV and $b_{max}$ = 3 Å, out of 6000 trajectories using classical friction, 4234 are trapped and 1766 are not trapped. Among the ground state components of untrapped trajectories, 1653 lead to NRNCT and 113 lead to RNCT processes. Among the excited state components, 1493 lead to NRCT and 273 lead to RCT (reactants/products). In case of semiclassical friction, the numbers are 1301 (trapping), 3505 (NRNCT), 1194 (RNCT), 3994 (NRCT) and 705 (RCT), respectively. If no sign correction (quantum friction) is involved, the numbers are 70 (trapping), 4157 (NRNCT), 1773 (RNCT), 3577 (NRCT) and 2353 (RCT), respectively.

For $E_{coll}$ = 5.0 eV and $b_{max}$ = 3 Å, out of 6000 trajectories using classical friction, only 578 trajectories are trapped. Among the ground state components of untrapped trajectories, 5059 lead to NRNCT and 363 lead to RNCT processes. Among the excited state components, 4390 lead to NRCT and 1032 lead to RCT (reactants/products). In case of semiclassical friction, the numbers are 16 (trapping), 4687 (NRNCT), 1297 (RNCT), 4775 (NRCT) and 1209 (RCT), respectively. If no sign correction (quantum friction) is involved, the numbers are 1 (trapping), 4615 (NRNCT), 1384 (RNCT), 4111 (NRCT) and 1888 (RCT), respectively.

To demonstrate that our classical trajectory calculations on the 2-state PES lead to meaningful results, we compare them with the earlier reported quantum dynamical results, trajectory surface hopping results and the available experimental results in Table 1 as well as in Figure 4.

Table 1: Integral cross-section values obtained from two electronic state classical trajectory calculations for $b_{max} = 3$ Å in comparison with earlier quantum dynamics results [57], TSH results [34] and experimental findings [49].

| | type | NRNCT | RNCT | NRCT | RCT | Trap | Quantum Dynamics [57] | TSH [34] | Expt. [49] |
|---|---|---|---|---|---|---|---|---|---|
| 1.7 eV | classical | 0.49 | 0.12 | 0.29 | 0.0 | 6.96 | NRNCT: 4.13, RNCT: 3.77, NRCT: 0.0, RCT: 0.0 | NRNCT: 2.38, RNCT: 3.69, NRCT: 0.04, RCT: 0.15 | --- |
| | semi class | 2.23 | 1.53 | 2.30 | 0.0 | 3.53 | | | |
| | quantum | 5.35 | 2.79 | 0.20 | 0.76 | 0.53 | | | |
| 2.5 eV | classical | 0.56 | 0.15 | 0.35 | 0.0 | 4.37 | NRNCT: 2.62, RNCT: 1.94, NRCT: 0.24, RCT: 0.27 | NRNCT: 1.72, RNCT: 1.88, NRCT: 0.12, RCT: 0.24 | RNCT: 2.04, RCT: 0.27 |
| | semiclassical | 2.26 | 1.62 | 2.40 | 0.0 | 0.92 | | | |
| | quantum | 3.29 | 1.95 | 0.14 | 0.34 | 0.13 | | | |
| 5.0 eV | classical | 0.96 | 0.48 | 0.81 | 0.0 | 1.29 | NRNCT: 1.06, RNCT: 0.74, NRCT: 0.07, RCT: 0.53 | NRNCT: 0.57, RNCT: 0.85, NRCT: 0.32, RCT: 0.13 | |
| | semiclassical | 1.54 | 0.99 | 1.53 | 0.0 | 0.10 | | | |
| | quantum | 1.48 | 0.98 | 0.07 | 0.05 | 0.36 | | | |

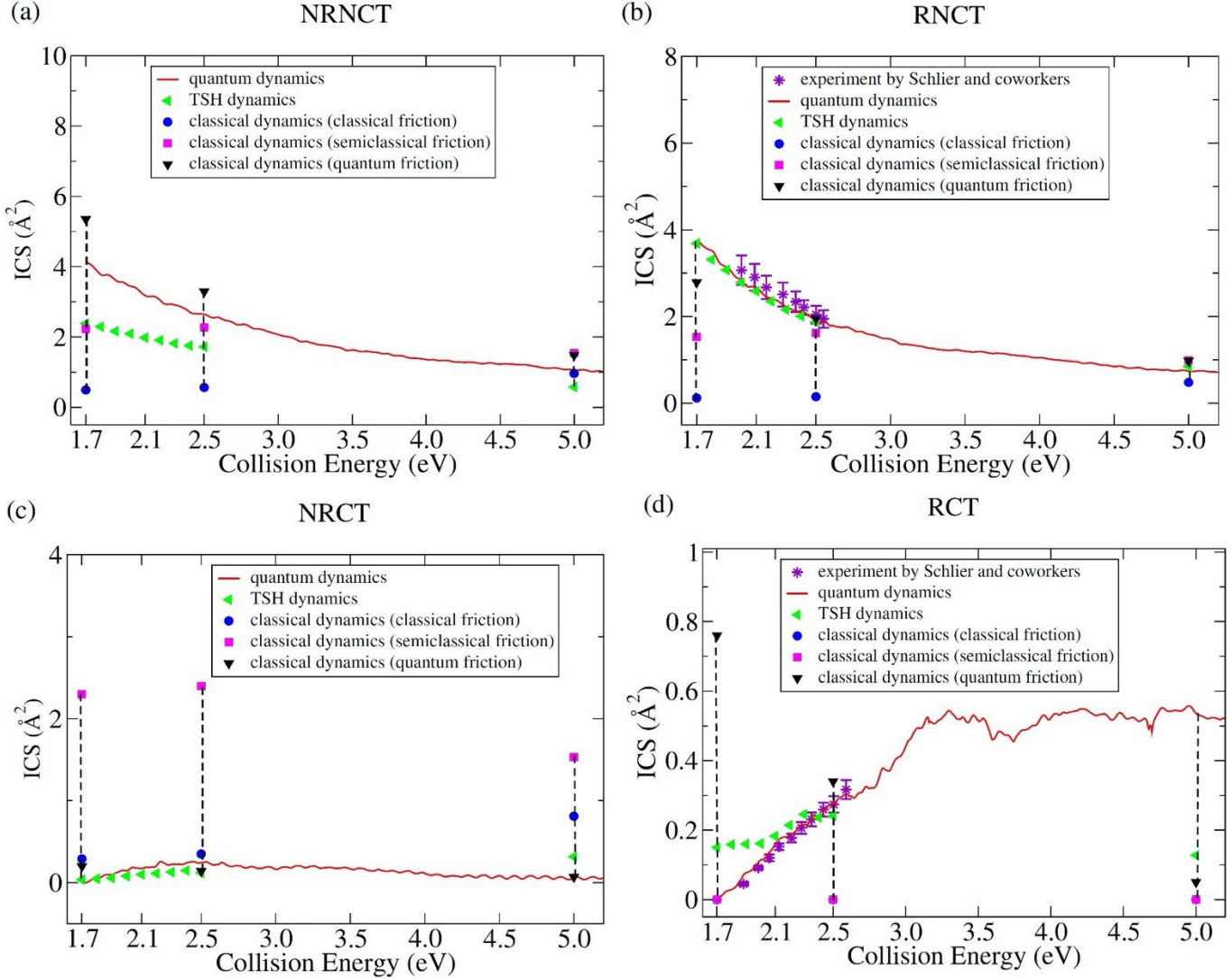

Figure 4: Comparison of the computed cross section results (using classical, semiclassical and quantum friction, $b_{max}$ = 3 Å) with the earlier reported quantum dynamical results [57], TSH results [34] and the available experimental data [49] for NRNCT, RNCT, NRCT and RCT processes.

We had demonstrated earlier [43] that NACTs could be treated as friction in a classical mechanical framework and that the dynamics on a single adiabatic PES with friction showed trapping cross section decreasing from 1.2 Å² at $E_{coll}$ = 1.7 eV to 0.4 Å² at $E_{coll}$ = 2.5 eV. The same set of classical trajectory calculations with friction showed the cross section for RNCT to be about 3 Å² at $E_{coll}$ = 2.5 eV, in agreement with quantum dynamical result (1.94 Å²), TSH result (1.88 Å²) and experimental result (2.04 Å²).

In the present paper, using $b_{max}$ = 3 Å (as used in our earlier study), we obtain the cross section for the RNCT process at $E_{coll}$ = 2.5 eV to be 0.15 Å² using 2-state calculations with classical friction, 1.62 Å² with semiclassical friction, and 1.95 Å² with quantum friction. Under the same conditions, the cross section for RCT is found to be 0.0 Å² (classical friction), 0.0 Å² (semiclassical friction) and 0.34 Å² (quantum friction) when compared to 0.27 Å² from quantum dynamics [57], 0.24 Å² from TSH [34] calculations, and 0.27 Å² from expt.[49]. The cross section for trapping is found to be 4.37 Å² (classical friction), 0.92 Å² (semiclassical friction) and 0.13 Å² (quantum friction). The cross section for trapping decreases from 6.96 Å² at $E_{coll}$ = 1.7 eV to 4.37 Å² at $E_{coll}$ = 2.5 eV to 1.29 Å² at 5.0 eV (classical friction); from 3.53 Å² to 0.92 Å² to 0.10 Å² (semiclassical friction); from 0.53 Å² to 0.13 Å² to 0.36 Å² (quantum friction). It is important to add that the present results using quantum friction agree well with the quantum dynamical results at higher energies ($E_{coll}$ = 5.0 eV) also. While the cross section for the NRCT process is overestimated by the quantum friction model, that for the RCT process is underestimated when compared to the exact quantum results over the entire range.

## 5. Summary and Conclusion

By considering the quantum mechanical NACTs as equivalent to friction in classical mechanics, we have studied the ($D^+$, $H_2$) collision dynamics on a 2-state PES and shown that there could be substantial slowing down of the dynamics (due to nonadiabatic coupling) resulting in the formation of the triatomic species $DH_2^+$ as well as RNCT, RCT and NRCT processes. The 2-state classical trajectory calculations with quantum friction yield reliable results when compared to quantum and experimental results at $E_{coll}$ = 2.5 eV. The use of semiclassical friction also yields comparable results. The use of classical friction, however, seems to exaggerate the cross section for trapping and underestimate the cross section for the RNCT process. Unfortunately, there are no experimental results available to compare with, at this point in time.

**Acknowledgement:** This study was supported by the SERB grant CRG/2019/000793 to Satrajit Adhikari. One of us (NS) is grateful to the Indian National Science Academy for the INSA Distinguished Professorship.

**Conflict of interest**: The authors have no conflicting interests.

# Appendix: Formulation of Integral Cross Sections For Classical Trajectory Calculation

While employing Quantum Dynamics, the integral cross-sections are calculated using the following expression:

$$\sigma = \frac{\pi}{\kappa_{vj}^2} \sum_{v'=0}^{v'_{max}} \sum_{j'=0}^{j'_{max}} \sum_{J=0}^{J_{max}} (2J+1) P^J_{v'j' \leftarrow vj}$$

In classical calculations, the probability for a specific final roto-vibrational state, $P^J_{v'j' \leftarrow vj}$ becomes $\sum_{N_{v'j'}} (1/N_T)$, where $1/N_T$ represents the probability for one specific trajectory leading to reactant/product in final roto-vibrational state $v'j'$ and $N_{v'j'}$ represents the total number of trajectories leading to $v'j'$. Therefore,

$$\sigma = \frac{\pi}{\kappa_{vj}^2} \sum_{v'=0}^{v'_{max}} \sum_{j'=0}^{j'_{max}} \sum_{N_{v'j'}} (1 + 3 + \cdots + (2J_{max} + 1)) (1/N_T)$$

$$= \frac{\pi}{\kappa_{vj}^2} \sum_{v'=0}^{v'_{max}} \sum_{j'=0}^{j'_{max}} \sum_{N_{v'j'}} \frac{(J_{max}+1)}{2} \frac{(1 + 2J_{max} + 1)}{2} (1/N_T)$$

$$= \frac{\pi}{\kappa_{vj}^2} \sum_{v'=0}^{v'_{max}} \sum_{j'=0}^{j'_{max}} \sum_{N_{v'j'}} (J_{max}+1)^2 (1/N_T)$$

$$= \frac{\pi}{\kappa_{vj}^2} \sum_{v'=0}^{v'_{max}} \sum_{j'=0}^{j'_{max}} \sum_{N_{v'j'}} (l_{max}+1)^2 (1/N_T)$$

In classical limit $J \sim l$, see page 93 of the Book, "The Quantum Classical Theory" by Gert D. Billing [23]. Since the orbital angular momentum quantum number, $l_{max}$ is different for different trajectories, it is replaced by $l_{v'j'}$. Hence

$$\sigma = \frac{\pi}{\kappa_{vj}^2} \sum_{v'=0}^{v'_{max}} \sum_{j'=0}^{j'_{max}} \sum_{N_{v'j'}} (l_{v'j'}+1)^2 (1/N_T)$$

where $l_{v'j'} = l_{max} O_{v'j'}$, $O_{v'j'}$ is a pseudo random number.

On the other hand, as per Eq. 3.62 of the Book, "The Quantum Classical Theory" by Gert D. Billing [23], the impact parameter:

$$b = \frac{\hbar(l + 1/2)}{\mu v_0} \quad \rightarrow \quad l = \frac{b\mu v_0}{\hbar} - \frac{1}{2}$$

Therefore, for our case,

$$l_{v'j'} = l_{max}O_{v'j'} = \left(\frac{b_{max}\mu_{tri}v_0}{\hbar} - \frac{1}{2}\right)O_{v'j'}$$

where $b_{max}$, $\mu_{tri}$, and $v_0$ are the maximum impact parameter, triatomic center of mass and initial relative velocity, respectively.

[77] Maurer R.J.; Jiang B.; Guo H.; Tully J.C., Mode specific electronic friction in dissociative chemisorption on metal surfaces: H$_2$ on Ag (111). *Phys. Rev. Lett*. **2017**,118, 256001.

[78] Spiering P.; Shakouri K.; Behler J.; Kroes G.-J.; Meyer J., Orbital-dependent electronic friction significantly affects the description of reactive scattering of N$_2$ from Ru (0001). *JPCLett*. **2019**, 10, 2957- 2962.

[79] Maurer R.J.; Zhang Y.L.; Guo H.; Jiang B., Hot electron effects during reactive scattering of H$_2$ from Ag (111): assessing the sensitivity to initial conditions, coupling magnitude, and electronic temperature. *Faraday Discuss*. **2019**, 214, 105-121.

[80] Zhang Y.L.; Maurer R.J.; Guo H.; Jiang B., Hot-electron effects during reactive scattering of H$_2$ from Ag (111): the interplay between mode-specific electronic friction and the potential energy landscape. *Chem. Sci*. **2019**, 10, 1089-1097.

[81] Zhang Y.L.; Maurer R.J.; Jiang B., Symmetry-adapted high dimensional neural network representation of electronic friction tensor of adsorbates on metals. *J. Phys. Chem. C*, **2020**,124, 186-195.

[82] Auerbach, D. J.; Tully, J. C.; Wodtke, A.M., Chemical Dynamics from the gas-phase to surfaces, *Nat. Sci*., **2021**, e10005
28

# Supporting Information

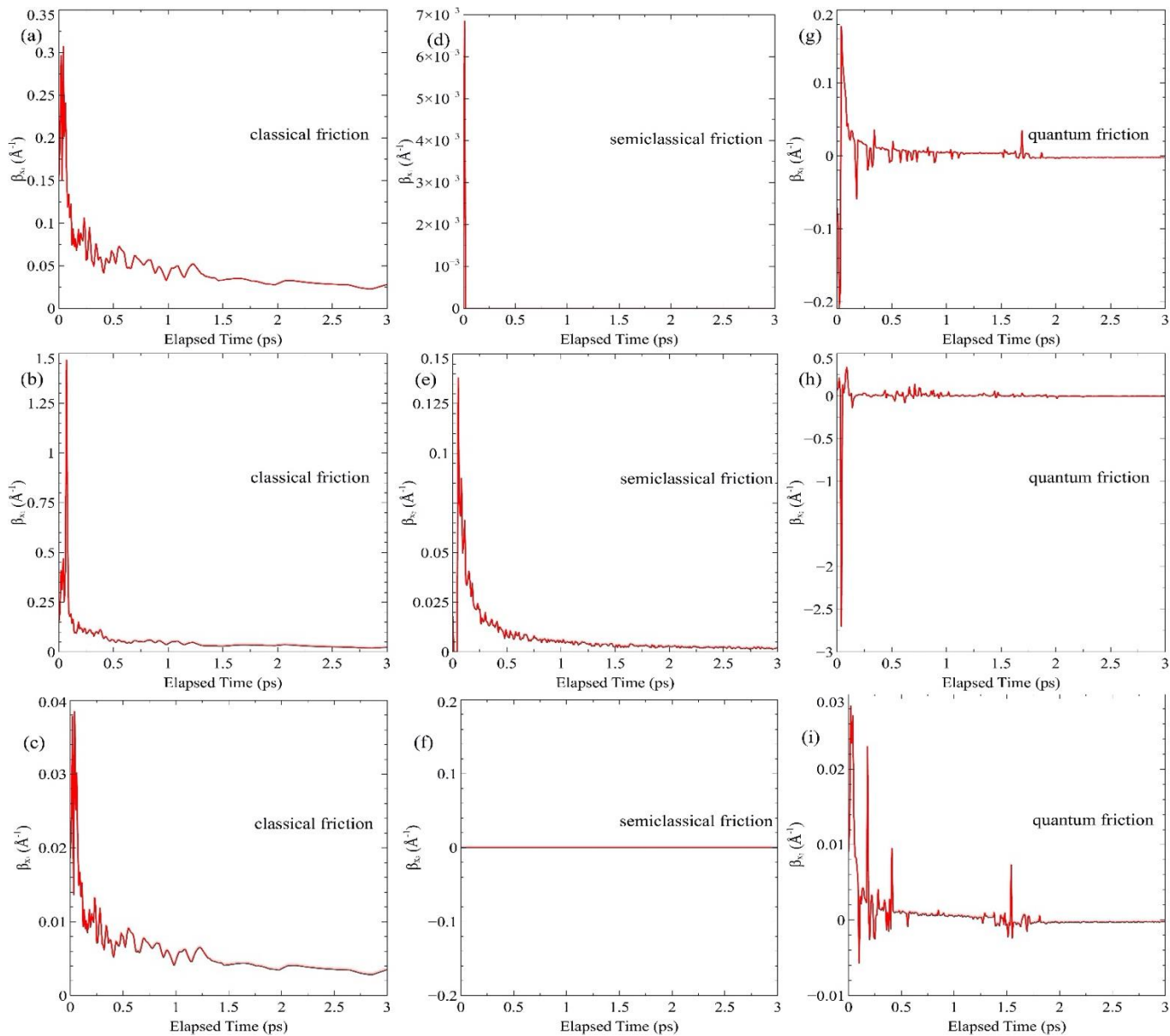

Figure S1: Variation of the *x*-component of the friction coefficient with time for a representative trajectory at $E_{coll}$ =1.7 eV ($b_{max}$ = 3 Å) for the ground electronic state of $DH_2^+$ (using three types of friction).

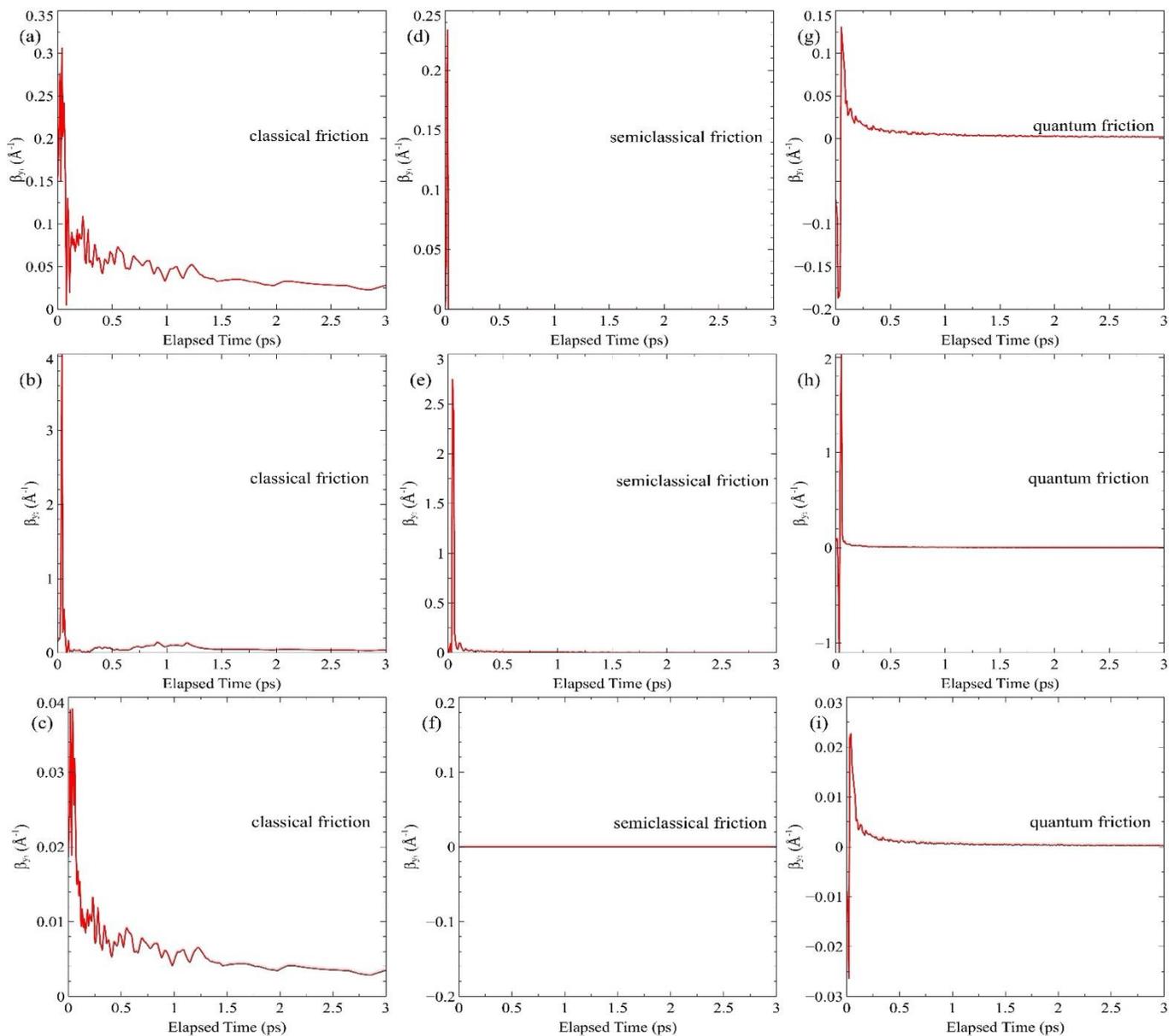

Figure S2: Variation of the *y*-component of the friction coefficient with time for a representative trajectory at $E_{coll}$ =1.7 eV ($b_{max}$ = 3 Å) for the ground electronic state of $DH_2^+$ (using three types of friction).

# **Graphical Abstract**

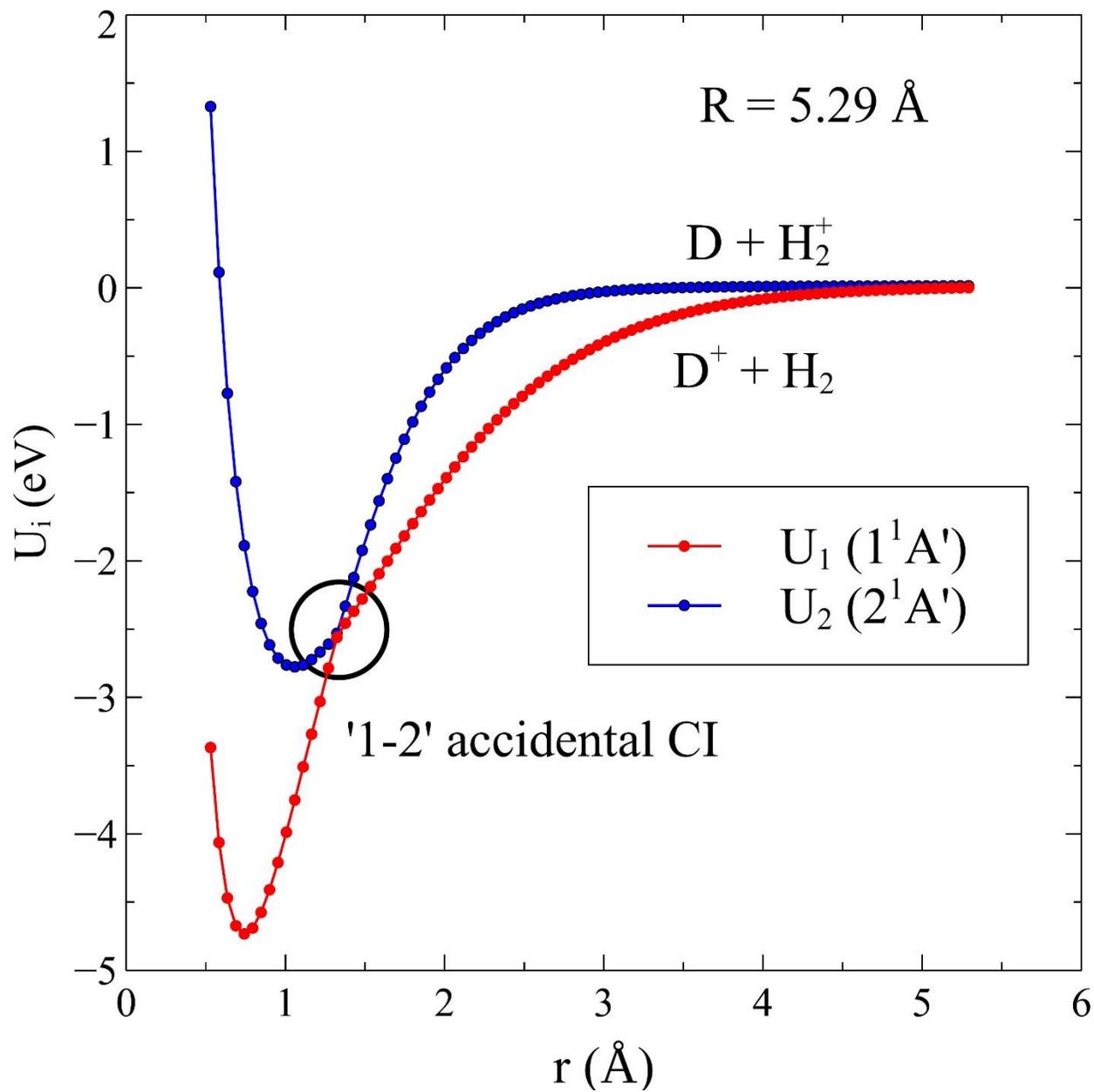